# The footprint of a metrics-based research evaluation system on Spain's philosophical scholarship: an analysis of researchers' perceptions

Ramón A. Feenstra[1]  & Emilio Delgado López-Cózar[2]

[1]Universitat Jaume I de Castelló (Spain) [2] Universidad de Granada (Spain)

**Abstract**

The use of bibliometric indicators in research evaluation has a series of complex impacts on academic inquiry. These systems have gradually spread into a wide range of locations and disciplines, including the humanities. The aim of the present study is to examine their effects as perceived by philosophy researchers in Spain, a country where bibliometric indicators have long been used to evaluate research. The study combines data from a self-administered questionnaire completed by 201 researchers and from 14 in-depth interviews with researchers selected according to their affiliation, professional category, gender and area of knowledge. Results show that the evaluation system is widely perceived to affect research behaviour in significant ways, particularly related to publication practices (document type and publication language), the transformation of research agendas and the neglect of teaching work, as well as increasing research misconduct and negatively affecting mental health. Although to a lesser extent, other consequences included increased research productivity and enhanced transparency and impartiality in academic selection processes.

Ramón A. Feenstra     Emilio Delgado López-Cózar
feenstra@uji.es     edelgado@ugr.es

# Introduction

The possibility of evaluating research performance using bibliometric indicators based on quantitative data from publications and citations emerged in parallel with the first citation indexes (Garfield 1963). Despite Eugene Garfield's warnings against the overuse of these metrics (Garfield 1979) and the debate on their foundations (Elkana 1978), they quickly became established (Narin 1976). For several decades now, research evaluation systems (RESs) based on bibliometric indicators have been implemented in many countries (Geuna and Martin, 2003; Whitley 2007), including Spain, Australia, UK, Slovakia, Poland, Sweden and China, to mention just a few (Butler 2003; 2007; Jiménez-Contreras et al. 2003; Hicks 2012; Hammarfelt and Rushforth 2017). This proliferation has spurred extensive debate about their possible effects (Wouters 2014; Wouters et al. 2015; Hicks et al. 2015; Aagaard, Bloch and Schneider 2015; Wilsdon et al. 2015; de Rijcke et al 2016).

One side of this debate points to effects such as the increase in publications and productivity (Jiménez-Contreras et al. 2003; Besir-Demir 2018)—although some authors question whether this is a direct consequence of the evaluation system (Osuna et al 2011)—and the internationalisation or promotion of research excellence (Hicks 2012; Van Den Besselaar, Heyman and Sandström 2017; Bautista-Puig, Moreno-Lorente and Sanz-Casado, 2020). Likewise, some studies find that metrics-based RES, in combination with other peer-review methods and mechanisms, can enhance the efficiency or transparency of evaluation processes (Butler 2007; Derrick and Pavone 2013).

On the other hand, authors have examined the potential effects on the research agenda, insofar as the plurality of research topics may be curtailed by the shift towards mainstream approaches and the neglect of more local issues (Butler 2007; Hicks 2012; Thelwall et al 2015; de Rijcke et al 2016; Giménez-Toledo 2016); or the pressure to publish high impact papers in English, leading to changes in publication practices, both in terms of document types (Hicks 2012; Wouters 2014) and the choice of publishing language (Ossenblok et al 2012; Hammarfelt and De Rijcke 2015; Hicks et al 2015). Similarly, highly competitive RESs can affect researchers' motives (Hangel and Schmidt-Pfister 2017), as their survival instinct drives them to consider the result (publication) as the end purpose and essence of their work (Butler 2007, Delgado-López-Cózar 2010, Wouters 2014; Rodríguez-Bravo and Nicholas 2018). Unsurprisingly, therefore, some studies observe that the pressure on researchers to publish has psychological repercussions, such as anxiety or depression (Díaz 1996; Gill 2009, Castañeda et al. 2014; Levecque et al. 2017; Canosa-Betés and de Liaño 2020), or neglect of other essential tasks such as teaching, tutoring or reviewing (Laudel and Gläser 2006; Wouters 2014; de Rijcke et al. 2016). Finally, a growing body of literature is examining how pressure to publish may be driving a rise in research misconduct (De Vries, Anderson and Martinson 2006; Delgado-López-Cózar, Torres-Salinas, and Roldán-López, 2007; John, Loewenstein, and Prelec, 2012; Pupovac, Prijić-Samaržija, and Petrovečki 2017; Felaefel et al. 2018; Maggio et al. 2019).

RESs vary greatly and their impact differs according to academic discipline (Hicks 2012; Aagaard, Bloch and Schneider 2015; Hinze et al. 2019; Nicholas et al. 2020b). Some studies find that the humanities are particularly vulnerable in this shift towards bibliometric systems due to their extensive and diverse range of disciplines (Laudel and Gläser 2006; Hicks 2012; Thelwall, et al. 2015; Hammarfelt 2017; Hammarfelt and Haddow 2018). In this line, quantitative measures are regarded as particularly ineffective in the humanities because of specific characteristics such as their more pronounced national focus, higher use of vernacular languages, predilection for publishing books, single author approach, and higher number of publications addressed to a non-scholarly public (Hicks 2004; Nederhof 2006; Borrego and Urbano 2006; Ochsner, Hug and Daniel 2013). Indeed, some authors have suggested that the highly idiosyncratic nature of the humanities may be affected by the use of these evaluation systems (Hicks 2012; Hicks et al. 2015; Hammarfelt 2017). In consequence, an extensive literature is now examining, for example, the specific impact of RESs on the types of document published in the humanities (Ossenblok et al. 2012; Hammarfelt and De Rijcke 2015; Giménez-Toledo 2016; Guns and Engels 2016; Hammarfelt 2017; Engels et al. 2018).

The impact of RESs on research has raised concerns at an international level. Defined by some authors as the "metric tide" (Wilsdon et al. 2015; Wouters et al. 2015), the responsible use of metrics has been urged, notably, in the Declaration on Research Assessment (DORA, 2012) and the Leiden Manifesto (2014). Calls have also been made for research into the impact of RESs in specific settings and countries (Wouters 2014; Hammarfelt and De Rijcke 2015; Hinze et al. 2019).

In the case of Spain, studies have examined the characteristics of the evaluation model and its impacts on science (Rey-Rocha et al. 2001; Jiménez-Contreras et al. 2002; Jiménez-Contreras et al. 2003; Butler 2004; Fernández, Pérez and Merchán 2006; Osuna et al. 2011; Delgado-López-Cózar 2010; Fernández, Díaz and Ramos 2011; Molas-Gallart 2012; Hicks 2012; Derrick and Pavone 2013; Marini 2018; Rodríguez-Bravo and Nicholas 2018; Bautista-Puig, Moreno-Lorente and Sanz-Casado 2020). Scholars have also analysed the effect of the Spanish RES on the social and human sciences (Ibarra, Castro and Barrenechea 2007; Giménez-Toledo 2016), or have critically examined the demands of the system in its application in the humanities (Rovira 2006; Rodríguez-Yunta 2017). Specific research has analysed the fields of, for example, economics and business studies (Delgado and Fernández-Llera 2012; Cancelo and Bastida 2013), communications (Rodríguez and Gil-Soldevilla 2018), philology (Urbano et al. 2004), translation (Granell Zafra 2015), history (Cañibano et al. 2018) and archaeology (Canosa-Betés and de Liaño 2020). Other research has gathered data from scientists, including those working in the humanities, on their impressions and perceptions of the Spanish RES (Buela-Casal and Sierra 2007; Buela-Casal 2007; Giménez-Toledo 2016).

In the area of philosophy, several contemporary studies have denounced the implementation of the 'neoliberal university model', widely regarded as a threat to critical thinking and intellectual diversity (Sinead 2019). However, to the best of our knowledge no research has explored perceptions of the effect RESs are having on one of the humanities' core disciplines: philosophy. The present study attempts to fill this gap by examining the issue in the Spanish context. Spain is an ideal test-bed, as it has been at the forefront of the metric tide for several decades now (Delgado-López-Cózar, Ruiz-Pérez and Jiménez-Contreras, 2007). Analysing the current impact on Spanish philosophy may alert us to trends that could be spreading in other contexts.

# Spain

As in other countries, Spain's RES is explicitly determined by research performance and metrics (Hicks 2012; Derrick and Pavone 2013). In the 1980s, the first RESs were introduced to measure output performance reflected mainly in publications. Their objective was to open up Spanish science to the wider world, increase research productivity and introduce evaluation mechanisms and competition for resources. Rules were implemented to regulate access to and promotion in scientific careers, and to control the distribution of resources and academic status (Sanz-Menéndez 1995; Fernández, Pérez and Merchán 2006; Cruz-Castro and Sanz-Menéndez 2007; Delgado-López-Cózar 2010; Fernández, Díaz and Ramos 2011).

Crucial to the adoption of bibliometric indicators was the creation, in 1989, of a research incentive system known as the *sexenio*. This incentive, administered by the National Commission for the Evaluation of Research Activity (CNEAI), assessed each researcher's individual performance every six years, and if successful, they were rewarded with a financial bonus (Jiménez-Contreras et al. 2003). Researchers had to present five publications, which were evaluated according to two basic quality indicators:

1. The relevance and prestige of the journal or publisher that had published the article or book, with the stipulation that in disciplines with established international quality criteria for publications, these criteria will prevail. In all knowledge areas except the humanities and law, "preference will be given to research published in recognised prestigious journals, which are defined as those occupying relevant positions in the lists, by scientific areas, in the subject category listing of the Journal Citation Reports of the Science Citation Index (Institute of Scientific Information, Philadelphia, PA, USA). If none of the categories in these listings coincides with the applicant's specialism, the commission will draw up an appropriate list, classifying the journals that best represent the specialism according to their impact index" (official state gazette (BOE), 1996; our translation).
2. The bibliographical citations of the work that reflect its impact.

These general criteria remain unaltered to this day, making Spain a prime example of consistent use of the same evaluation policy, based on bibliometric indicators, over time.

The 2001 Law on Universities took this culture of evaluation a step further by establishing the National Agency for Quality Assessment and Accreditation (ANECA). This agency is responsible for regulating access and promotion of teaching staff in higher education. The creation of the ANECA confirmed the significant weight of research in general, and of publications in particular, in the process of acreditation (review process for promotion) (Table 1). It also confirms the importance of the *sexenios* by directly awarding 15 points for each *sexenio* obtained (out of 80 point needed for been acreditated as full professor), and attributing the highest possible score to teaching staff with four *sexenios*. The evaluation regulations introduced by the ANECA use the same wording to refer to publication evaluation as the CNEAI, although some specific details were added according to knowledge areas (ANECA 2007; 2008). Since 2006, 130,685 academics have been evaluated by the agency, of whom 79,112 obtained the accreditation (60.5%) (ANECA 2019a).

**Table 1.** Minimum thresholds to obtain accreditation for tenure-track and tenure positions in Spain, and maximum weight given to publications within the scoring system (as published by ANECA).

| Tenure-track and tenure positions | Scores (out of 100) | |
|---|---|---|
| | Minimum threshold | Weight of publications in scoring system |
| Lecturer | 55 | 42 |
| Senior Lecturer | 65 | 38 |
| Professor | 80 | 35 |

Since then, the *sexenio* has become increasingly important in the academic research and teaching career. It is now an essential requirement for supervising doctoral theses, endorsing scholarships, teaching official masters and PhD programmes, leading research projects and participating in academic examination committees (Delgado-López-Cózar and Martín-Martín 2019). Since 2012, it has even been used to increase the teaching load of academics without *sexenios* and to reduce that of colleagues that do have them (BOE 2012). The evaluation agencies of some Spanish regions and universities also calculate the complementary bonuses they award on the basis of *sexenios* (ANECA 2010). In sum, *sexenios* are used to differentiate and stratify university teaching staff (Delgado-López-Cózar 2016). The implication is, therefore, that although evaluation is purportedly voluntary, over the years it has become an unavoidable obligation. While almost 50,000 academics were evaluated between 1989 and 2007 (Agraït and Poves 2009), the equivalent number in the four-year period 2016–2020 was 39,000 (CNEAI 2020). At the same time, researchers have increasingly adapted to this evaluation system, as reflected in the rising success rates: in 2007, 35.8% of academics did not have a *sexenio* (Agraït and Poves 2009), in 2013 this percentage had fallen to 27.1% (Ministry of Education, Culture and Sport, 2014) and in 2019, it had dropped again to 21.8% (Ministry of Universities, 2020). However, it is also true that in 2019 fewer than half (45.8%) of university teaching staff had all the *sexenios* they could have accumulated over their academic careers.

The application of bibliometric indicators—especially journal rankings—in the social sciences and humanities came up against a problem, however: the lack of Spanish publications that met the reference standard (Journal Citation Reports) set by the CNEAI and ANECA. In an attempt to solve this problem, numerous journal evaluation platforms have appeared, including DICE (dissemination and editorial quality of Spanish humanities, social science and legal journals)[1], RESH (Spanish social science and humanities journals)[2], IN-RECS, IN-RECJ, IN-RECH (impact index for social science, law and humanities journals, respectively)[3], CARHUS Plus+[4], MIAR (information matrix for journal analysis)[5], CIRC (integrated scientific journals classification)[6], visibility and impact ranking of humanities and social science journals with the FECYT quality certificate[7], and H Index for Spanish scientific journals according to Google Scholar Metrics (2007–2018)[8].

Although slight modifications have been introduced into the journal evaluation criteria over the last thirty years and the sources used in evaluation have changed (Ruiz-Pérez et al. 2010), the presence of bibliometric indicators continues to grow steadily in the humanities.

The analysis of the CNEAI evaluation criteria for the humanities, including philosophy, in recent years confirms the growing weight and full assimilation of bibliometric indicators (Marini, 2018). Current *sexenio* requirements in the area of philosophy recognise as key quality indexes "the inclusion of international databases such as the Arts and Humanities Citation Index of the Web of Science, Journal Citation Reports, Social Sciences Edition, Emerging Sources Citation Index and Scimago Journal Rank" (BOE 2019, our translation). The agency has recently published its scoring system table for the first time (see Table 2).

**Table 2.** Guidelines applied by the CNEAI's 2019 assessment of research sexenios. Area of Philosophy, Philology and Linguistics (ANECA 2021)

| Articles | | | |
|---|---|---|---|
| JCR/SCOPUS Q1-Q3 ICDS: 9+ | ESCI SCOPUS Q4 CARHUS B ICDS: 7-8 | CIRC ERIH PLUS CARHUS C ICDS: 4-6 | Others CARHUS D ICDS: 0-3 |
| *Score*\* | | | |
| 8-10 | 7-8 | 6-7 | 2-6 |
| **Books** | | | |
| SPI Spanish publishers 1-30\*\* Foreign publishers 1-33\*\* IE-CSIC (high) | SPI Spanish publishers 31-46 Foreign publishers 34-40 IE-CSIC (medium) | SPI Spanish publishers 47-50 Foreign publishers 41-43 IE-CSIC (medium) | SPI Spanish publishers 51-55 Foreign publishers 44-48 IE-CSIC (low) |
| *Score* | | | |
| 9-10 | 7-8 | 6-7 | 2-6 |
| **Book chapters** | | | |
| SPI Spanish publishers 1-30 Foreign publishers 1-33 IE-CSIC (high) | SPI Spanish publishers 31-46 Foreign publishers 34-40 IE-CSIC (medium) | SPI Spanish publishers 47-50 Foreign publishers 41-43 IE-CSIC (medium) | SPI Spanish publishers 51-55 Foreign publishers 44-48 IE-CSIC (low) |
| *Score* | | | |
| 7-8 | 7 | 6-7 | 2-6 |

ICDS: Secondary Composite Index Broadcasting (https://miar.ub.edu/about-icds)
CIRC: Integrated Scientific Journals Classification (https://www.clasificacioncirc.es)
SPI: Scholarly Publishers Indicators (http://ilia.cchs.csic.es/SPI)
CARHUS: Classification of Social Science and Humanities Journals (https://agaur.gencat.cat/es/avaluacio/carhus)
SCOPUS = meaning Scimago Journal & Country Rank SJR https://www.scimagojr.com/journalrank.php
IE-CSIC: Editorial index CSIC (https://www.cid.csic.es/biblioteca/node/113)

\* Scale: 0 to 10 points
\*\* Publisher position in the ranking

In turn, the promotion accreditation system (ANECA) conclusively enshrined the use of bibliometric criteria with the approval of Royal Decree 415/2015 (BOE 2015), the requirements of which were introduced in 2017. The criteria have been updated since then and today, the requirements set out in the following table (Table 3) must be met to obtain a full professorship in the areas of philosophy and ethics.

Table 3. Summary based on the evaluation criteria approved in 2019 and applied as of 15 January 2021, update of 2017 criteria (ANECA 2019b; ANECA 2019c)

| Philosophy (Arts and Humanities) | Philosophy Moral (Social Sciences) |
|---|---|
| "present at least 45 publications. In the case of scientific articles, half of them must have been published in first level journals. At least two must have been published in category Q1 of the Web of Science (WoS), and a further two in category Q2." | "present at least 50 publications (indexed articles or otherwise, books and book chapters), of which at least 30 should be journal articles". "At least 16 articles should have been published in level 1 journals (80% for full professorship category A)" |

| **Definition of journal levels** | |
|---|---|
| Level 1: <br><br> a) Web of Science (WoS) databases: Sciences Citation Index (JCR), Social Sciences Citation Index (JCR). Journals only included in the Emerging Sources Citation Index (ESCI) are excluded. <br><br> b) Q1 and Q2 in the Scimago Journal Rank (SJR). <br><br> c) Q1 and Q2 in the ranking of journals with the FECYT quality certificate. <br><br> Level 2: <br><br> a) Indexed journals in SRJ Q3 and Q4. <br><br> b) Journals classified in Q3 and Q4 of the ranking of journals with the FECYT quality certificate. | Level 1 journals: JCR Q1, Q2; SJR/ Scopus Q1 <br><br> Level 2 journals: JCR Q3 and Q4; SJR/Scopus Q2 and Q3; Dialnet Métricas Q1; and FECYT Q1/Q2 |

These criteria show that publishing in high impact journals is essential to career advancement in philosophy, and in the humanities in general. In this study we examine Spanish university researchers' perceptions of this phenomenon.

## Methodology

The study combines a survey and 14 in-depth interviews in order to gather both quantitative and qualitative data. Discovering the impact of RES is a complex task, in which a mixed-methods approach is particularly appropriate for examining specific contexts (Butler 2010; Hammarfelt and De Rijcke 2015). The quantitative data are invaluable for analysing issues such as the perception of RES impact on publication practices (formats and languages), while the qualitative data provide deeper insights into the impacts of the Spanish RES as perceived by the researchers.

*Self-administered questionnaire*

The study population comprised university researchers working in the knowledge areas of philosophy and ethics in Spain. We identified the members of this academic community through a systematic search of the websites of Spanish universities and research centres. In the vast majority of institutions we were able to identify affiliated members and researchers, with only four exceptions. Through these inquiries we identified 541 faculty members and researchers, of whom 521 worked in universities and 20 in the research centre CSIC; 44 universities (37 public and seven private) took part in the study and responses were received from all but three institutions. Table 4 shows the distribution by knowledge area.

**Table 4.** Demographics of the survey of Spanish university faculty and researchers in philosophy and ethics

| FIELD | Total population | Respondents | Average response rate |
|---|---|---|---|
| Philosophy | 380 | 115 | 30.5% |
| Ethics | 161 | 86 | 52.8% |
| Total | 541 | 201 | 37.1% |

The data was collected using Google Forms and respondents were asked about their affiliation and employment situation. The online survey remained open for responses between February and June 2019. On 25 February, a message was sent to the institutional email address of the 541 faculty members and researchers identified, followed by two reminders, two weeks and four weeks after the initial contact. We also approached the main scientific societies and associations for Spanish philosophy professionals—the Spanish Association for Ethics and Political Philosophy (AEEFP), the Academic Society of Philosophy (SAF) and the Spanish Philosophy Network (REF)—requesting their collaboration. In May, these organisations contacted their members by email to encourage them to participate in the survey. The survey was closed on 14 June 2019.

The survey included the following questions:

P.1 To what extent do you consider that the rise and fall of ethics and philosophy journal ratings in their corresponding bibliometric indicators affects your research career? (determining, strong, moderate, weak, very weak, none, do not know and/ or answer)

P.2 Are you satisfied with the importance given to publications and the way they are evaluated by the CNEAI (*sexenios*)? (very satisfied, satisfied, indifferent, dissatisfied, very dissatisfied)

P.3 Are you satisfied with the importance given to publications and the way they are evaluated by the ANECA (accreditations)? (very satisfied, satisfied, indifferent, dissatisfied, very dissatisfied)

P4 Which of the following document types (books, articles, book chapters, conference papers, new digital media) do you think is the most appropriate for publishing your research results (1 not at all appropriate, 5 very appropriate)

P.5 Which language do you think is the most suitable for disseminating research in your knowledge area? (choose more than one option if appropriate)

The questionnaire also included an open question where respondents could freely express their opinions on the object of study. The 60 responses we obtained to this question provided qualitative data in the survey stage of the research.

## Interviews

The interviews took place in September and October 2019. The 14 interviewees were selected according to the criteria of affiliation, professional category, gender and disciplinary area in order to guarantee the widest possible range of profiles. Seven of the interviewees were men and seven were women, and seven worked in the field of ethics and seven in philosophy (Table 5).

**Table 5.** List of university faculty and researchers in philosophy and ethics interviewed according to career position.

| Nº | Career position | Area | Code |
|----|-----------------|------|------|
| 1 | Research fellow | E | RfE1-i |
| 2 | Research fellow | E | RfE2-i |
| 3 | Research fellow | Ph | RfPh1-i |
| 4 | Lecturer | Ph | LPh1-i |
| 5 | Lecturer | Ph | LPh2-i |
| 6 | Senior lecturer | Ph | SlPh1-i |
| 7 | Senior lecturer | Ph | SlPh2-i |
| 8 | Senior lecturer | Ph | SlPh3-i |
| 9 | Senior lecturer | E | SlE1-i |
| 10 | Senior lecturer | E | SlE2-i |
| 11 | Professor | Ph | PPh1-i |
| 12 | Professor | E | PE1-i |
| 13 | Professor | E | PE2-i |
| 14 | Professor | E | PE3-i |

The interviewees were affiliated to the universities of Barcelona, Castellón, Complutense (Madrid), Granada, Murcia, Valencia, Zaragoza and the Basque Country, as well as the Institute of Philosophy at the CSIC. The semi-structured interviews lasted an average of 35.30 minutes; the shortest was 14.14 minutes and the longest, 59.10 minutes. The interviews were then transcribed for analysis. The questions were posed in an open-ended format. Specifically, a generic question was asked about the positive and negative effects that researchers associate with the RES.

**Table 6.** List of acronyms

| Career stage | |
|---|---|
| Research Fellow | Rf |
| Post-doctoral researcher | Pdr |
| Lecturer | L |
| Senior Lecturer | Sl |
| Professor | P |
| **Area** | |
| Ethics | E |
| Philosophy | Ph |
| **Source of information** | |
| Interview | -i |
| Open section of the survey | -s |

## Results

The first part of the study aimed to determine whether the researchers perceived journal bibliometric indicators (e.g., JCR, SJR) as affecting their professional careers (Q.1). The responses are strikingly convincing: 80% of Spanish university philosophy and ethics teaching staff consider that their careers are conditioned by the bibliometric trends of the journal in which their articles were published; 56% consider them to be decisive or very decisive, and just 9% only a little or not at all decisive (Figure 1).

**Figure 1.** Opinions of university philosophy and ethics faculty and researchers in Spanish institutions about the effect on their professional career of the rise or fall of ethics and philosophy journals in their corresponding bibliometric indicators

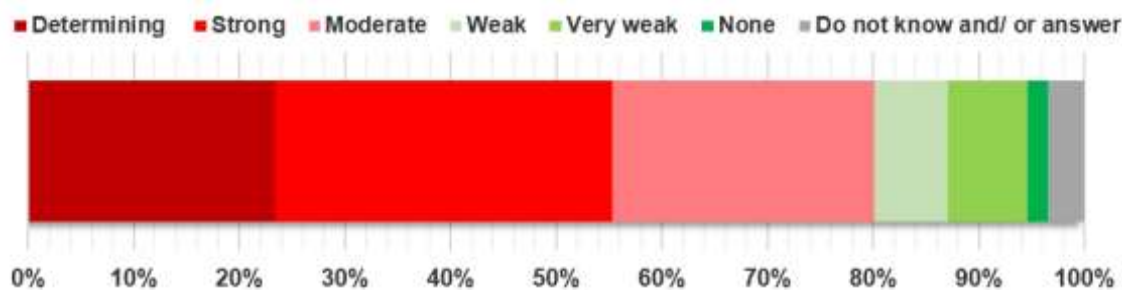

Their level of satisfaction with the way the ANECA and the CNEAI evaluate publications (Q.2 and Q.3) is also notable. Here, we observe a general dissatisfaction, although levels of discontent are much higher in the case of the ANECA (Figure 2): 57.7% of those surveyed claimed to be dissatisfied or very dissatisfied, compared to just 27.3% who are satisfied or very satisfied. The degree of dissatisfaction with the CNEAI is lower: 43.8% are dissatisfied or very dissatisfied, although this figure is much higher than the percentage who claimed to be satisfied or very satisfied (31.1%). These data are interesting, precisely because the ANECA attributes greater weight to bibliometric indicators in their research evaluations, as we saw in section 2. These results are similar to findings from other studies that also revealed researchers' dissatisfaction (around 51%) with the Spanish RES (Sanz-Menéndez and Cruz-Castro 2019).

**Figure 2**. Level of satisfaction of university philosophy and ethics faculty and researchers in Spanish institutions with the weight the CNEAI and the ANECA give to publications and the way they evaluate them

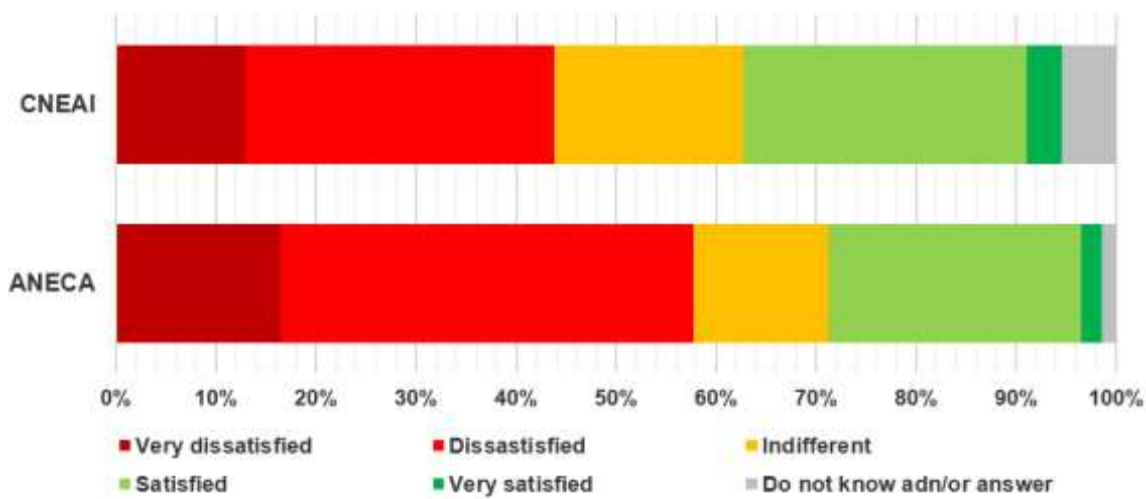

The researchers' responses explicitly pointed to a direct relationship between the Spanish RES and their research work. They identified a wide range of effects, some negative and others positive. The harmful effects they perceived mainly concerned modifications in their publication practices (more limited range of document types and publication languages), changing research agendas, neglect of teaching and other tasks, growing research misconduct in publishing, and damaging psychological effects. The positive effects, although mentioned less frequently, included stimulating productivity and increased transparency and impartiality in academic selection processes. The respondents varied in the importance they attributed to the effects presented in the survey, however. In the next section we examine the researchers' opinions on the most outstanding effects. We have included as many opinions as possible, given the high level of expositional and explanatory clarity of the qualitative material.

## Enhanced transparency and impartiality in academic selection processes

RESs are designed to be simple to apply, and to have clear criteria and the potential to encourage transparency (Butler 2007; Derrick and Pavone 2013). Indeed, one positive effect the researchers perceived is the system's ability to tackle what is defined as the 'patronage' model. They consider that the current system establishes public criteria for impartial evaluation. Some researchers stated that the RES has finally put a stop to a model in which nepotism and arbitrariness determined the fate of academic careers. This idea was mentioned by eight interviewees at different stages in their careers (RfE1-i, SlPh1-i, SlE1-i, SlE2-i SlPh3-i, PE3-i, PE2-i, PPh1-i) and by one person in the open section of the questionnaire (PdrPh1-s). We reproduce some of the most pertinent verbatim extracts below:

> it's true that now researchers are accredited by an impartial institution. SlE2-I

> I've known academics who have never passed an accreditation, and they are in the university for reasons that I don't fully understand. The fact that there is a filter that stops departments from hiring just anyone is positive I think, because it will eventually put an end to something that has functioned in Spain in a very grievous way, and that is the almost militaristic nature of [university] departments, with very deep-rooted hierarchies. I know the generation that came before mine, who entered the university under the wing of a renowned professor but had no previous accreditations, and you can see much higher levels of subservience. This has brought about an independence, a value, a minimum guarantee. That's the value of the agencies, that they have set criteria to determine whether a person meets the minimum research requirements to receive a bonus in return. I think it's a good thing that the agencies exist. SlPh3-i

The ANECA has an advantage, on the one hand, and that is that it publishes the accreditation criteria you are evaluated on. And this has gone some way to eliminating the nepotism in universities. RfE1-i fortunately university institutions have implemented mechanisms that make it much more difficult for this 'mandarinate' relationship to thrive like it used to. Put another way, your PhD supervisor or your department head or chair doesn't necessarily have to like you, because if you are good and you get accredited (you'll be accredited outside your department) and they normally have automatic mechanisms to create positions for accredited academics. So, what used to happen, that you had to convince whoever was the full professor at that time to create your position, and he had the final word, that's no longer the case because one way or another your professional career is decided somewhere different from your place of work […] All this has helped to tone down, to moderate the vertical hierarchical relationships… they likely still exist, but there are now incentives that might mean they will disappear. SlPl1-i

Some minimum guidelines are certainly desirable. And it's true that a structure gives an orientation and it makes sense to have some filters in place. PE3-i

One respondent went a step further, defining the Spanish metrics-based evaluation system as an attempt to "break the lock of corruption, nepotism and cronyism" (PdrPh1-s). Another two academics claimed to be in tenured positions thanks to the national centralised external evaluation system.

The positive thing for me is that I am now a senior lecturer. I'm sure that if it hadn't existed (ANECA or a similar type of agency), I wouldn't be in the university or I'd be on precarious contracts SlE1-i

I'm working in the public university system thanks to ANECA. I was in a private university, and if it hadn't been for the national acreditation process I probably wouldn't have got into the public system. I'm personally very grateful [….] Because in this case, what accreditations have achieved is that academic promotion doesn't depend on whatever hierarchy happens to be in place at the time. SlPh1-i

Another beneficial effect of the RES is that it has enabled a higher proportion of women to enter the Spanish university system (PE2-i; RfE1-i). A final observation came from a full professor who cautioned that although the problem of impartiality has declined, we should bear in mind that the negative effects of the system (which we examine below) are deep-seated and affect the way research is carried out in philosophy.

I hear a lot of people saying, "I got promotion thanks to the ANECA, because the chair [of my department] didn't let me, etc." There are some people who consider having an external evaluation to be positive compared to the local "fiefdoms". Agreed. But the truth is that it's had some very perverse effects on the way we do research. PPh1-i

## Stimulation of research and productivity

Several of the researchers noted the incentive to do research and the prevention of complacency as positive consequences of the system. Three senior lecturers and a professor made this point in the interviews (SlPh1-i, SlPh2-i, SlPh3-i, PE2-i) and a lecturer raised the same issue in the open question of the survey (LE1-s). Some of the most pertinent comments were:

[the *sexenios*] force people to keep active and struggle a bit harder [...] So having this stimulus is no bad thing. I don't know if people would work in the same way without it, I think things would get a bit out of control... SlPh2-i

[the agencies] have made people pull their socks up. SlPh1-i

We have to recognise that our academia had plenty of shirkers, people who would get tenure and then slink off to sleep without ever doing anything again. PE2-i

These researchers thought the system was fulfilling one of its main objectives: to stimulate research (Hicks 2012; Van Den Besselaar, Heyman and Sandström 2017). Recall that the *sexenios* were introduced in 1989 precisely with this aim in mind (see section 2).

## Changes in preferred document types for publication

Responses to the question on researchers' preferred document types (Q.4) for publication revealed a clear inclination towards books and articles (Figure 3). Notably, these two document types were ranked as equally relevant, followed by book chapters, which were also highly regarded.

**Figure 3. Most suitable document types for publication according to philosophy and ethics faculty and researchers in Spanish institutions**

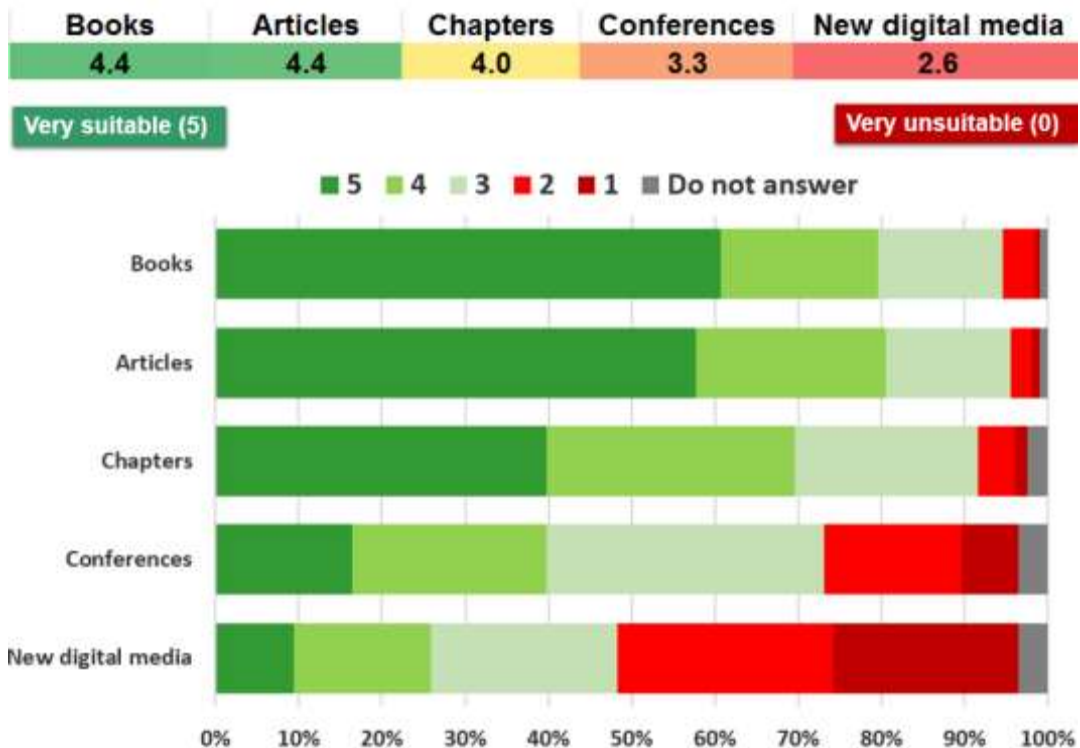

However, the qualitative data contribute some interesting nuances to the survey data, showing that, in general, philosophy researchers value book publication differently to articles. Although some consider that all publication channels are important and no one channel should be prioritised over others (SlPh1-i, SlE2-i, PE3-1), there is a certain agreement that books as a vehicle for publication are typical of and specific to philosophy, an opinion held especially by tenured lecturers and professors with permanent contracts (SlE5-s PPh3-s, PE2-I, SlPh3-I, PPh1-i, SlPh2-i).

There does seem to be a widespread consensus, however, that the evaluation agencies regard books with a certain disdain. Coinciding with other contributions (Hammarfelf and Rijcke 2015; Hammarfelf 2017; Engels et al. 2018), the interviewees' responses showed that research evaluation criteria condition publication behaviour in Spain. Several argued that the system's prioritisation of journal articles had restricted the plurality of document types. The following extracts are illustrative:

> In recent years articles have been prioritised over single-author books. In general, books are underrated by the evaluation agencies [….]. It's affected me personally. In one of the *sexenios* I applied for, one of the books I published [….] scored less than an article. Strange. I don't understand why this type of publication must be valued less than articles in indexed journals. There's no academic justification for it [….]. It's forcing all young researchers to focus their activity on publishing research articles in indexed journals. PPh1-i

> I find it incomprehensible that publishing an article in a particular journal is sometimes rated more highly than an entire book published by a good publisher. SlE5-s

We're losing dissemination through books and a more contemplative way of thinking. First because we don't have enough time to sit down and write, and second, because of ANECA demands. We know we'll score practically the same for publishing an article as we would for a book, which are totally incomparable in terms of the work involved. So people are saying no to publishing books. I see it among my colleagues, that once they are settled and they don't need to build up their curriculum, that's when they spend more time writing books. But when you're engaged in the process necessary for accreditation, you go for what takes least time. SlE1-i

Put simply, the way research evaluation is carried out in the field of philosophy is dreadful. We can adapt: instead of racking my brains over what book I'm going to write, I turn out ten-page articles with an introduction and conclusions and with nothing to say, and that's that. But it's no good for research in philosophy. SlPh3-I

In an area like ethics and political reflection, the connection with the general public gets lost in the current scientific demands, especially publishing in English in international journals. A large part of the practical sense of normative or critical reflection is lost if it doesn't reach society, and this is what's happening with the growing push towards scientistic models typical of other disciplines. The loss of value of the book as a benchmark of scientific output is the clearest example of this. SlE4-s

There are, however, differences between younger researchers and their established colleagues. The early career researchers are pragmatic and adapt to the rules of a system that rewards papers with a high impact factor. Although in general this trend does not affect the humanities as much as other disciplines (Nicholas et al. 2020a; Nicholas et al. 2020b), it has been reported in previous studies (Jamali et al. 2020). One young researcher expressed a preference for the book format in the open question of the survey (LE4-s), but most interviewees said they preferred articles (RfPh1-i, RfE1-i, LPh1-i), essentially because it is much more 'cost effective' than writing a book (it calls for less effort, the publication process is quicker and it gives a better payoff for their careers). The following verbatim extract is illustrative on this point:

With the rush to keep up in publishing requirements, articles are much easier and more attainable. It's much easier in the sense that it's faster than writing a book […] and you get feedback from people more quickly… from the scientific community you belong to. What's more, it's much simpler to read an article than a book […] it can be cited more quickly and it has a faster impact than a book, and you can link them up more quickly and widely […] LPh1-i

## Preferred languages for publishing

The preference in RESs for high impact papers can also affect the languages used in publishing (Hammarfelt and De Rijcke 2015; Hicks et. al 2015; Ossenblok et al. 2012). Our survey (Q.5) found that a clear majority of philosophy academics in Spain (70%) consider that ideally, philosophical inquiry is best expressed through more than one language (Figure 4), with Spanish being the preferred language (mentioned by 87.6% of the respondents), followed by English (72.1%), and at a considerable distance, German. In fact, most of the researchers considered two languages to be essential (48%), with Spanish and English being the most popular combination. The multilingualism reflected in these data is a cognitive characteristic typical in this field, as is the option to use the vernacular language as the main communication medium. The recent growth of English signals its use as the lingua franca in the scientific community.

**Figure 4.** Most appropriate languages for publishing according to university philosophy and ethics faculty and researchers in Spanish institutions

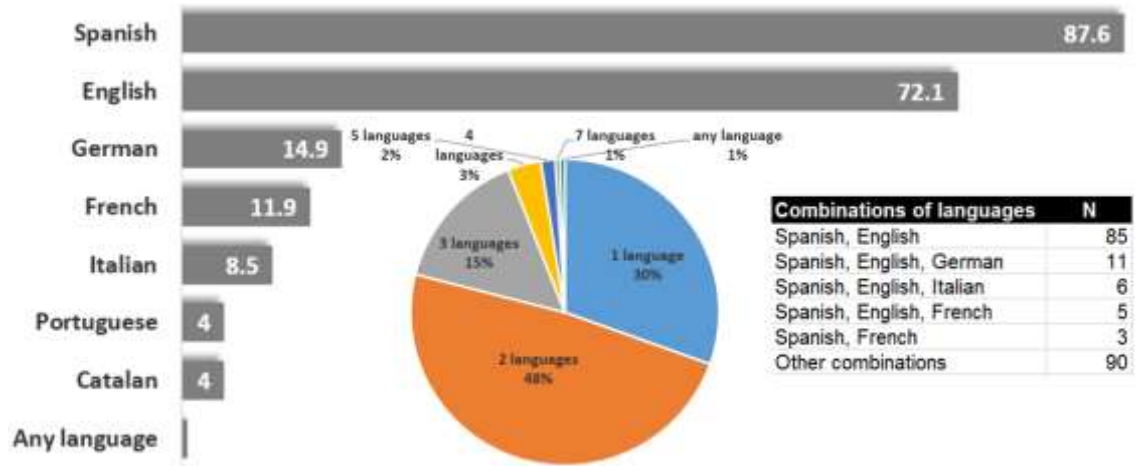

In the qualitative part of the study, again the results were more nuanced; the RES was criticised for having promoted the use of English to the detriment of other languages. The researchers regard the language used in philosophy as a key element bound up with the culture and with different philosophical traditions. Although to a lesser extent than the criticisms voiced on document types, this issue was also raised in the open section of the survey (SlE1-s, SlE4-s, SlPh5-s, SlPh11-s, PPh1-s, PPh5-s) and in the interviews (SlPh3-i). We observed a certain polarisation against the growing dominance of English, and at the same time, defence of the use of Spanish. These attitudes are reflected in the following comments:

> With regard to the evaluation criteria, it's my view that in the humanities (and particularly in philosophy), language is not neutral. In the end, we are developing culture, humanism, with the aim of enriching our own cultural heritage and our own language. So this business of translating everything into English is a sign of imperialism. We're living in a world dominated by English culture. I understand that the lingua franca prevails in physics or chemistry or in medicine, but I don't understand it in philosophy, or in any other area of the humanities. I find it insidious, because one of the main criteria in the humanities is precisely the enrichment of one's own culture. SlPh3-i

> The imposition of English as the language of scientific publishing is intolerable and very damaging to science and research freedom… PPh1-s

> The bias towards the English-speaking world to the detriment of Latin, Germanic, Italian, French... [cultures] has done great harm to philosophy and to research. The whole system needs changing… PPh5-s

> The parallel with scientific journals only in English does great harm. French, German or Italian are not valued to the same extent. SlPh5-s

> Excessive importance of English… SlE1-s

> In an area like ethics and political reflection, the connection with the general public gets lost in the current scientific demands, especially publishing in English in international journals... SlE4-s

> It is vital to take care of philosophy produced in Spanish… SlPh11-s

Other participants expressed less polarised views, recognising the positive aspects as well as pointing out some difficulties associated with the spread of English:

> […] it's essential to write in English, but we don't write like native speakers and that's a hurdle to acceptance. It's frustrating. SlE6-s

> The problem is that if you really want to enter into dialogue you have to do it in English. That has its good and bad sides. The good thing is that you can't use much rhetoric and you communicate with more people, and your ideas have to be clearer. The bad thing is that you lose obvious expressive capacities that are also part of philosophy. […] And some expressivity is lost, but to continue advancing knowledge (so to speak), they [journals published in English] are an ideal medium in academia and you learn a lot from them. SlPH2-i

> […] I know there are problems when you submit articles to certain journals that can tell you aren't a native English speaker and that's where there are numerous difficulties, for example. RfPh1-i

## Transformation in research agendas

Another impact identified in the study, particularly in the interviews, is the possible modification of research agendas. The researchers considered that the current model stifles plurality in what is essentially a heterogenic field of knowledge. They understand that one type of issue or specific philosophical perspective is encouraged to the detriment of others (LPh2-i, SlPh2-i, SlE2-i, PPh1-i, PE2-i). The potential impact on intellectual diversity (Butler 2007; Hicks 2012; Thelwall et al 2015; De Ricjke et al 2016) appears to be another consequence of the RES. Various comments back up this idea:

> It makes no sense for philosophy, because in addition only one line is being prioritised, and that's the analytical line, they're defining their position with their Q1 journals there. And the most fundamental aspect of philosophy is being sidelined. LPh2-i

> Well, because the agencies have decided to evaluate philosophy in these terms. According to this type of journal, in the field of philosophy it is easier to publish in a top-level journal if you're working in analytical philosophy. If you aren't, it's much more complex (as in my case). You have to play the game with somebody else's rules. SlPh2-i

> I think that if things continue as they are, the subjects we work with will eventually be transformed […] But beware of adopting the forms of science and scientific production imposed by the publishers. SlE2-i

The risk of centring all philosophical output in academic papers is that they end up acting as a mechanism for censoring and restricting that philosophical diversity and creativity of thought. I think articles are a good thing. I myself write rigorous, serious articles (from a historiographic point of view) on an author or a concept, but I think this is just one part or aspect of philosophical production. PPh1-i what we're seeing in the humanities and the social sciences is that they are applying criteria that are closely related to the world of the hard sciences and also of some social sciences that I call "recalcitrant analytics". This lobby of analyticists and technicians behind it, who have grown up in this belief, who I call the "cuckoos" (Q1, Q2, etc.), thinking that this is the best. That is, bringing into a field of knowledge evaluation systems that do not match that field. PE2-i

As part of this concern, the interviewees also referred to the gradual abandonment of research as something they do with passion and interest. Several interviewees consider that research today is simply the pursuit of immediate academic survival, a view also found in other studies (Rodríguez-Bravo and Nicholas 2018). This idea has taken root in particular among early career researchers, who described how they adapt to the demands of the system. The means—publishing—has become an end in itself, a prediction already made in earlier research (Butler 2007; Wouters 2014; Hangel and Schmidt-Pfister 2017; Delgado-López-Cózar and Martín-Martín 2019). On this point, the philosophy academics claim that their prime concern is to publish in the way that most efficiently advances their academic career. The following extracts illustrate these reflections:

People no longer do research on what it is important to study; they study what they think will allow them to move forward in their academic careers [...] Maybe we are romantics, but one thinks there should be an emotional link with the subject of one's inquiry [...]. One understands that certain problems affect one personally. That is what I would define as intellectual heteronomy. One is willing to work on what is demanded of one. The only thing I want is to be guaranteed some possibility of success. SlPh1-i

My PhD experience. I was obviously worried about what would happen afterwards. But I felt a sense of calm working on my thesis that they don't have today. They [young PhD students] are more concerned about where they can publish and how they can build their curriculum than the progress of their doctoral thesis. And this is an experience I share with my colleagues and that they understand perfectly. I can't blame my doctorands for having this professional attitude when they will actually end up on the street if they don't take this path and operate in this way. SlPh3-i short-termism, wanting to publish quickly and have a high impact on top of that [...] some research lines get lost in favour of other more productive ones. In other words, the subjects studied aren't the most relevant, or scientifically and socially interesting, but topics that allow us to progress academically. SlE2-i

[...] it has this type of effect: a custom-designed curriculum. That means it's not a curriculum people build up because they are interested in the research topic, but because they keep an eye on the calls and they design it in line with the calls. PE3-i

What I'm seeing in younger people, PhD students, brilliant individuals that one comes across, is that they follow designer research and scientific career paths. They are no longer thinking so much about doing good research or providing good results, but what they are thinking is where to submit their paper so it will have this impact, so it counts towards [their career], etc. The way researchers work has changed dramatically, and greatly so in our area of humanities where the way of working is different from the hard sciences. I notice a lack of innovation and creativity precisely because of the constraints imposed by this type of model, which at the same time leaves behind those who do not want to submit to it. So if someone wants to continue doing research they must know how to adapt. But as you adapt, you are also to some extent forgoing what could be your own contribution, the part you can develop, and I think that's a shame. PE1-i

## Proliferation of research misconducts

A specific study of research misconduct in Spain found that 91.5% of academics in the areas of ethics and philosophy considered research misconduct to be on the rise. The following types of misconduct stood out in particular: duplicate publication, self-plagiarism, the use of personal influence and citation manipulation; 90.5% of the researchers associated the spread of research misconducts with the RES applied in Spain (Authors, 2021).

## Appearance of mental health problems

As in previous research (Díaz 1996; Gill 2009; Castañeda et al 2014; Levecque et al 2017; Goyanes and Rodríguez 2018; Canosa-Betés and de Liaño 2020), several interviewees in our study referred directly to the stress caused by the pressure to meet the demands of the system. Notably, these problems arise at all stages of the academic career, not only among early career researchers. This emotional impact is described by seven of the interviewees (RfPh1-i, LPh1-i, SlPh2-i, SlPh3-i, SlE1-i SlE2-i, PE3-i).

It's not the amount of work that I find stressful. It's the lack of expectations. The fact of having done so many things and having no guarantee… and being totally exposed to the inability to make plans for more than one year ahead in my life. RfPh1-i

[…] And as for the criteria specifically, they are increasingly more complicated and what causes anxiety is that it never ends. The career path (or obstacle race) is interminable. LPh1-i

There's a kind of stress. And that's strange because now we have the figure of the stressed philosopher, which is a contradiction […] they stimulate an often unnecessary 'hyperproduction'. It's good and bad at the same time. The stimulation is good, but it can be experienced as stifling […] SlPh2-i

[…] it's leading to anxiety, an awful lot. Even at a professional level, no longer just among interns, it's beginning to be a reason for people losing interest in research. I know more and more academics who give up and just teach their classes, that's all. They lose their interest and motivation for philosophy. SlPh3-i

> Damaging, unbearable, this can turn into a pathology because when a person is being asked to lead goodness knows how many research projects, have recognised teaching in countless areas, publications in different areas of impact, teaching, research and, on top of that, administrative responsibilities, pedagogic innovation… they are required to do everything. Everything, and all with a level of excellence […] What isn't reasonable is to demand all three, because there isn't time for it all. And obviously that leads to health problems, anxiety, not being able to lead a normal life. SlE2-i

> People are anxious and the only thing they are interested in is publishing in journals that meet certain conditions. PE3-i

It is also striking to hear various senior lecturers openly stating that the present criteria put them off considering future promotions (SlE1-i, SlPh2-i, SlPh3-i). In some cases they even openly contemplate giving up research and the chance to advance in their academic careers. The following extracts illustrate these ideas:

> I don't think I would be accredited now for a senior lecturer position. It's a lot of Q1 level articles. I've never even considered going for a full professorship, I ruled that out from the start. It's very sad… but I've ruled it out. […] I want to publish in a different way, publish a book now. Yes, I'll try to get more *sexenios*, but at a different pace. In some way this marks your career and your objectives… That feeling that there is something external, that superego that's constantly telling you what you have to do and that stops you from enjoying [your work] in another way. SlE1-i

> It isn't a path that I'm attracted to right now, despite the economic motivation. In my department some people have decided not to struggle to publish or to meet the criteria, and they just teach their classes. SlPh2-i

> I know more and more academics who withdraw, they teach their classes and that's all. They lose their interest and motivation for philosophy. I know more and more people who seek refuge in their teaching, or do their research on the sidelines of academia, who don't worry about the next promotion or getting a positive evaluation, but what's important to them is their personal work and what they believe. That is, people who are becoming critical of the system and are working independently. I think this is a phenomenon, that all of us wonder if it makes any sense to continue playing the game, or whether we should give up and do what we enjoy, even if it means giving up any chance of promotion…. SlPh3-i

## Undervaluing of teaching: load and abandonment

A final possible effect of RESs often identified is the neglect of other important tasks, especially those related to teaching (Laudel and Gläser 2006; Wouters 2014; Rijcke et al. 2016). Some interviewees suggested that this is a consequence of the system, although to a much lesser extent than other issues (PE1-i, PPh1-i, LPh2-i, RfPh1-i). Their interpretations vary slightly depending on the stage they have reached in their careers.

> Since research takes up a lot of time, it seems that teaching is pushed into the background, and teaching well and providing good materials on Moodle aren't counted as valuable in the evaluations, only the papers you publish, etc. This means that it is also leading to a departure from the function of the university […]. I've heard about colleagues' dilemmas who've said "either I teach my classes well, or I do research", and that is terrible. PE1-i

What you do is pressurise university teachers to give up other kinds of activities, so they neglect other activities that have a major social and academic function and focus all their activity on publishing articles in indexed journals. This has a very damaging effect on the university system. PPh1-i

I see that people have a heavy teaching load (I'm not saying that they shouldn't have any teaching), and most of them pass it on to people in a much more vulnerable situation, instead of it being done by people in more stable positions and who don't have problems of professional or economic stability. RfPh1-i

## Discussion and conclusions

The main finding of this study is the deep, multilayered impact the Spanish RES has on academic work in philosophy. Although the researchers in our study identified some positive effects, such as increased productivity and more transparent policies in the academic promotion process, most of its consequences were deemed negative, notably its impact on research agendas, document types, publication language and mental health.

Although other studies involving early career researchers find that, on an international scale, metrics seem to have a lower impact on the humanities than in other fields (Nicholas et al. 2020a; Nicholas et al. 2020b), our study suggests that it leaves a deep mark on philosophy research in Spain. In general, our results seem to coincide with the effects analysed in other studies that warn of the potential impact metrics can have in the humanities (Whitley 2007; Hicks 2012; Thelwall et al. 2015; Hammarfelt 2017; Hammarfelt and Haddow 2018). Additionally, they confirm what has been widely theorised and examined, namely, that the evaluation system shapes the day-to-day dynamics of researchers' lives (Hicks 2012; Wouters 2014, Wouters et al. 2015; Hicks et. al. 2015; Aagaard, Bloch and Schneider 2015; Wilsdon et al. 2015; De Rijcke et al. 2016; Nicholas et al. 2020). Spanish researchers, especially early career researchers, adapt to the environment as a way of surviving in a highly competitive landscape, confirming the results of previous studies (Rodríguez-Bravo and Nicholas 2018; Jamali et al. 2020).

This article uses the actors' narratives to explore these consequences through their deeply illustrative and significant responses to the survey questions and in the interviews. They clearly describe how they consider not so much the end pursued in their research, but the means through which to communicate it. This is not surprising given that, in Spain, academic careers depend on the impact of the medium that publishes their results. The researchers describe how this causes a kind of alienation, a loss of the sense of scientific identity and of belonging to the field of knowledge that they contribute to, leading to an abandonment of the research ideal that guides or informs the scientific ethos. Passion for cognition is replaced by the desire for recognition, that is, concern that their research will be effective in advancing their scientific career. In sum, the pragmatic motives noted in some studies (e.g., Hangel and Schmidt-Pfister 2017) seem to trump epistemic or personal ones.

Future studies could extend this research through empirical analysis of each impact examined in this paper, for example, by studying the impact on the research project agenda over the last ten years, or on the number of books published by philosophers in Spain. The present study set out to explore the impact in as holistic a way possible in order to include the whole range of effects as perceived by actors in two specific knowledge areas in Spain: philosophy and ethics. We believe the results offer some interesting revelations.

Previous research, such as that of Hammarfelt and De Rijcke, warn of the difficulties of making causal claims between the implementation of evaluation models and their effects, while concluding that it is still "too early to talk about a 'metric culture' in the humanities" (Hammarfelt and De Rijcke 2015, 70). However, their study also provided clear evidence of the changes brought about by RESs in publications in the humanities for the context of Uppsala University. Our results point in the same direction and even suggest that the metric culture is penetrating a newly configured DNA among philosophers. The information presented here shows a keen awareness of the impact metrics are having in their fields and, at least in Spain, it is not unusual to hear philosophers talking at great length about bibliometric rankings, impact factors and journal quartiles. More research is clearly needed to corroborate the extent of the metric culture in the humanities, but this study is a wake-up call to the consequences of the implementation and sustained development of RESs in areas of the humanities such as philosophy.

## Ethical statement

Participants of the survey and the interviews were informed about the nature of the study and that their anonymity would be preserved. Verbal informed consent was obtained prior to each of the interviews. A summary of the data gathered during the research was published in an open report previously endorsed by by the three main Spanish philosophy and ethics associations: AEEFP, SAF and REF.

## Acknowledgements

The authors would like to thank Daniel Pallarés-Domínguez for his help in the data collection process. We would also like to thank the three main Spanish philosophy and ethics associations, AEEFP, SAF and REF, for their collaboration during the research process and their endorsement of the data collection report.

## Notes

[1] See http://epuc.cchs.csic.es/dice/
[2] See http://epuc.cchs.csic.es/resh/que_es
[3] See https://web.archive.org/web/20140713064650/http://ec3.ugr.es/in-recs/, https://web.archive.org/web/20140711021904/http://ec3.ugr.es/in-recj//, https://web.archive.org/web/20140713064650/http://ec3.ugr.es/in-rech/
[4] See https://agaur.gencat.cat/es/avaluacio/carhus/
[5] See http://miar.ub.edu/
[6] See https://clasificacioncirc.es/inicio
[7] See https://calidadrevistas.fecyt.es/ranking
[8] See http://hdl.handle.net/10481/57716

# References

Aagaard, K., Bloch, C., and Schneider, J. W. (2015) 'Impacts of performance-based research funding systems: The case of the Norwegian Publication Indicator' *Research Evaluation*, *24*/2: 106-117. doi: 10.1093/reseval/rvv003.

Agraït, N., and Poves, A. (2009) Informe sobre los resultados de las evaluaciones de la CNEAI. Available at:
http://www.iuma.ulpgc.es/~nunez/doctorado0910/CNEAISexeniosInfo2009v5.pdf

ANECA (2007) 'Programa de evaluación de profesorado para la contratación. Principios y orientaciones para la aplicación de los criterios de evaluación'. Madrid: ANECA.
http://www.aneca.es/content/download/11202/122982/file/pep_criterios_070515.pdf

ANECA (2008) 'Programa ACADEMIA. Principios y orientaciones para la aplicación de los criterios de evaluación'. Madrid: ANECA.
https://web.archive.org/web/20160206095755/http://www.aneca.es/content/download/10527/118089/version/1/file/academia_14_ppiosyorientaciones.pdf

ANECA (2010) 'Informe sobre el estado de la evaluación externa de la calidad en las universidades españolas en 2009'. Madrid: ANECA
http://deva.aac.es/include/files/deva/informes/evaluacion_externa/Informe_Calidad_2009.pdf?v=202012120360

ANECA (2019a) 'Memoria de activities'. Madrid: ANECA.
http://www.aneca.es/content/download/15516/190756/file/memoria_2019.pdf

ANECA (2019b) 'Méritos evaluables para la acreditación nacional para el acceso a los cuerpos docentes universitarios. Artes y Humanidades'. Madrid: ANECA.
http://www.aneca.es/content/download/15229/187663/file/Criterios%20Academia%202020b_Arte%20y%20Humanidades.pdf

ANECA (2019c) 'Méritos evaluables para la acreditación nacional para el acceso a los cuerpos docentes universitarios. Ciencias Sociales y Jurídicas y Humanidades'. Madrid: España.
http://www.aneca.es/content/download/15232/187693/file/Criterios%20Academia%202020_CC%20Sociales%20y%20Jur%C3%ADdicas_04.pdf

ANECA (2021). Orientaciones aplicadas por los comités evaluadores en la convocatoria 2019 de sexenios de investigación CNEAI. Madrid: ANECA.
http://www.aneca.es/Sala-de-prensa/Noticias/2021/Orientaciones-aplicadas-por-los-comites-evaluadores-en-la-convocatoria-2019-de-sexenios-de-investigacion

BOE (1996) Ministerio de Educación y Cultura, 20/11/1996, p. 35027-35032
https://www.boe.es/eli/es/res/1996/11/06/(4)

BOE (2012) Jefatura del Estado, 21/04/2012, p. 30977-30984.
https://www.boe.es/diario_boe/txt.php?id=BOE-A-2012-5337

BOE (2015) Ministerio de Educación, Cultura y Deporte, 17/6/2015, p. 50319-50337
https://www.boe.es/buscar/doc.php?id=BOE-A-2015-6705

BOE (2019) Ministerio de Ciencia, Innovación y Universidades, 26/11/2019, p. 130004-130024. https://www.boe.es/eli/es/res/2019/11/12/(10)

BOE (2021) Ministerio de Universidades, 2/1/2021, p. 359-380.
https://www.boe.es/diario_boe/txt.php?id=BOE-A-2021-53

Bautista-Puig, N., Moreno Lorente, L., and Sanz-Casado, E. (2020) 'Proposed methodology for measuring the effectiveness of policies designed to further research', *Research Evaluation*. doi: 10.1093/reseval/rvaa021

Besir Demir, S. (2018) 'Pros and Cons of the New Financial Support Policy for Turkish Researchers', *Scientometrics*, 116: 1-16. DOI: 10.1007/s11192-018-2833-4

Borrego, Á., and Urbano, C. (2006) 'La evaluación de revistas científicas en Ciencias Sociales y Humanidades', *Información, cultura y sociedad*, 14: 11-27. doi: 10.34096/ics.i14.886

Buela-Casal, G. (2007) 'Reflexiones sobre el sistema de acreditación del profesorado funcionario de Universidad en España', *Psicothema*, 19/3: 473-482

Buela-Casal, G., and Sierra, J. C. (2007) 'Criterios, indicadores y estándares para la acreditación de Profesores Titulares y Catedráticos de Universidad', *Psicothema*, 19, 357-369.

Butler, L. (2003) 'Modifying publication practices in response to funding formulas', *Research evaluation*, *12*/1: 39-46. doi: 10.3152/147154403781776780

Butler, L. (2004) 'What happens when funding is linked to publication counts?' In *Handbook of quantitative science and technology research* (pp. 389-405). Springer: Dordrecht.

Butler, L. (2007) 'Assessing University Research: A Plea for a Balanced Approach', *Science and Public Policy*, 34/8: 565–74. doi: 10.3152/030234207X254404

Butler, L. (2010) 'Impacts of performance-based research funding systems: A review of the concerns and the evidence'. In Performance based Funding for Public Research in Tertiary Education Institutions: Workshop Proceedings OECD Publishing, Paris.

Cancelo Márquez, M., and Bastida Domínguez, M. (2013) 'La evaluación de la investigación en España: los sexenios en las áreas de economía y empresa CIRIEC-España', *Revista de Economía Pública, Social y Cooperativa*, 78: 265-292

Cañibano, C.,et al. (2018) 'The evaluation of research excellence and the dynamics of knowledge production in the humanities: The case of history in Spain'. *Science and Public Policy*, 45/6, 775-789. doi: 10.1093/scipol/scy025

Canosa-Betés, J., and de Liaño, G. D. (2020) 'La carrera investigadora en arqueología y su impacto en la salud mental de los investigadores predoctorales', *Complutum*, 31/2, 379-401. doi: 10.5209/cmpl.72490

Castañeda, M. et al. (2014) 'Las investigadoras de la Universidad Nacional Autónomo de México y los sistemas de evaluación'. En Norma Blazquez (ed.), *Evaluación académica: sesgos de género* (pp. 223-244) CEIICH-UNAM: Ciudad de México.

CNEAI. (2020) 'Sexenios de investigación. Convocatorias de 2017, 2018 y 2019'. Resultados a fecha de noviembre 2020. http://www.aneca.es/content/download/15679/192482/file/tablas%20sexenio%20investigaci%C3%B3n.xlsx

Cruz Castro, L and Sanz Menéndez, L. (2007) Research evaluation in transition. Individual versus Organisational Assessment in Spain. In: R. Whitley and J. Gläser (eds.). *The Changing Governance of Sciences* (205-223), Springer: Dordrecht.

De Vries, R., Anderson, M. S., and Martinson, B. C. (2006) 'Normal misbehavior: Scientists talk about the ethics of research', *Journal of Empirical Research on Human Research Ethics*, 1/1: 43-50. doi: 10.1525/jer.2006.1.1.43

Delgado, M. M., and Thelwall M. (2015). Arts and humanities research evaluation: no metrics please, just data. *Journal of Documentation*. Journal of Documentation, 71/4: 817-833 doi: 10.1108/JD-02-2015-002

Delgado, F. J., and Fernández-Llera, R. (2012) 'Sobre la evaluación del profesorado universitario (especial referencia a ciencias económicas y jurídicas'. *Revista española de documentación científica*, *35*/2: 361-375. doi: 10.3989/redc.2012.2.861

Delgado-López-Cózar, E. (2010) 'Claroscuros de la evaluación científica en España', *Medes: Medicina en Españo*l, 4: 25-29.

Delgado-López-Cózar, E. (2016) 'La universidad española en el diván'. In: Sacristán del Castillo, JA; Gutiérrez Fuentes, JA (eds), *Reflexiones sobre la ciencia en España: cómo salir del atolladero* (pp. 163-230). Fundación Lilly: Madrid.

Delgado-López-Cózar, E., Torres-Salinas, D. and Roldán-López, Á. (2007) 'El fraude en la ciencia: reflexiones a partir del caso Hwang', *El profesional de la información*, 16/2, 143-150. doi: 10.3145/epi.**2007**.mar.**07**

Delgado-López-Cózar, E., Ruiz-Pérez, R., and Jiménez-Contreras, E. (2007) 'Impact of the impact factor in Spain'. *British medical journal*, *334*(7593). doi: 10.1136/bmj.39142.454086

Delgado-López-Cózar, Emilio and Martín-Martín, Alberto (2019). 'El Factor de Impacto de las revistas científicas sigue siendo ese número que devora la ciencia española: ¿hasta cuándo?". *Anuario ThinkEPI'*, 13, e13e09. doi: 10.3145/thinkepi.2019.e13e09

Derrick, G. E., and Pavone, V. (2013) 'Democratising research evaluation: Achieving greater public engagement with bibliometrics-informed peer review' *Science and Public Policy*, *40*/5: 563-575. doi: 10.1093/scipol/sct007

Díaz, Á. (1996) 'Los programas de evaluación en la comunidad de investigadores. Un estudio en la UNAM'. *Revista Mexicana de Investigación Educativa*, *1*/2: 408-423

Elkana, Y. (1978) 'Toward a metric of science: the advent of science indicators'. New York, NY: Wiley.

Engels, T. C., et al. (2018) 'Are book publications disappearing from scholarly communication in the social sciences and humanities?', *Aslib Journal of Information Management*, 70/6: 592-607, doi: 10.1108/AJIM-05-2018-0127

Felaefel, M,. et. al (2018) 'A cross-sectional survey study to assess prevalence and attitudes regarding research misconduct among investigators in the Middle East', *Journal of academic ethics, 16*(1), 71-87, doi: 10.1007/s10805-017-9295-9

Fernández Esquinas, M., Díaz Catalán, C., and Ramos Vielba, I. (2011) 'Evaluación y política científica en España: el origen y la implantación de las prácticas de evaluación científica en el sistema público de I+ D (1975-1994)'. In: González de la Fe, T., and López Peláez, A. (eds*): Innovación, conocimiento científico y cambio social: ensayos de sociología ibérica de la ciencia y la tecnología* (pp. 93-130). Centro de Investigaciones Sociológicas. Madrid.

Fernández Esquinas, M., Pérez Yruela, M., and Merchán Hernández, C. (2006) 'El sistema de incentivos y recompensas en la ciencia pública española. In: Sebastián, J. and Muñoz, E. (eds.). *Radiografía de la investigación pública en España* (p. 148-206). Biblioteca Nueva: Madrid.

Garfield, E. (1963) 'Citation indexes in sociological and historical research', *American documentation*, 14/4: 289-291. doi: 10.1002/asi.5090140405

Garfield, E. (1979) *Citation Indexing: Its Theory and Application in Science, Technology, and Humanitie*s.New York: John Wiley.

Geuna, A., and Martin, B. R. (2003) 'University research evaluation and funding: An international comparison', *Minerva*, *41*/4: 277-304. doi: 10.1023/B:MINE.0000005155.70870.bd

Gill, R. (2009) 'Breaking the Silence: The Hidden Injuries of Neo-liberal Academia'.I In Gill, R and Ryan-Flood, R (eds.), *Secrecy and Silence in the Research Process: Feminist reflections* (pp. 228–44). Routledge: London.

Giménez-Toledo, E. (2016) *El malestar de los investigadores ante su evaluación* Madrid:, Iberoamericana.

Goyanes, M., and Rodríguez-Gómez, E.F (2018) '¿Por qué publicamos? Prevalencia, motivaciones y consecuencias de publicar o perecer', *El Profesional de la Información*, 27(3), 548-558. doi: 10.3145/epi.2018.may.08

Granell Zafra, X. (2015). La evaluación de la investigación: criterios de evaluación en Humanidades y el caso de la Traducción e Interpretación. *Investigación bibliotecológica*, *29*/66: 57-78. doi: 10.1016/j.ibbai.2016.02.025

Guns, R., and Engels, T. C. (2016) 'Effects of performance-based research funding on publication patterns in the social sciences and humanities'. In *21st International Conference on Science and Technology Indicators-STI 2016. Book of Proceedings*.

Hammarfelt, B. (2017) 'Four claims on research assessment and metric use in the humanities', *Bulletin of the Association for Information Science and Technology*, 43/5: 33-38. doi: 10.1002/bul2.2017.1720430508

Hammarfelt, B. and De Rijcke, S. (2015) 'Accountability in Context: Effects of Research Evaluation Systems on Publication Practices, Disciplinary Norms and Individual Working Routines in the Faculty of Arts at Uppsala University', *Research Evaluation*, 24/1: 63–77. doi: 10.1093/reseval/rvu029

Hammarfelt, B., and Haddow, G. (2018) 'Conflicting measures and values: How humanities scholars in Australia and Sweden use and react to bibliometric indicators', *Journal of the Association for Information Science and Technology*, 69/7: 924-935. doi: 10.1002/asi.24043

Hammarfelt, B., and Rushforth, A. D. (2017) 'Indicators as judgment devices: An empirical study of citizen bibliometrics in research evaluation', *Research Evaluation*, 26/3: 169-180. doi: 10.1093/reseval/rvx018

Hangel, N., and Schmidt-Pfister, D. (2017) 'Why do you publish? On the tensions between generating scientific knowledge and publication pressure', *Aslib Journal of Information Management*, 69/5: 529-544. doi: 10.1108/AJIM-01-2017-0019

Hicks, D. (2004). The four literatures of social science. In: H. Moed (Ed.), Handbook of Quantitative Science and Technology Studies, Kluwer Academic Press, Dordrecht (2004), pp. 473-496

Hicks, D. (2012). Performance-based university research funding systems. *Research policy*, 41(2), 251-261. https://doi.org/10.1016/j.respol.2011.09.007.

Hicks, D., et al. (2015) 'Bibliometrics: the Leiden Manifesto for research metrics', *Nature*, 520/7548: 429-431. doi: 10.1038/520429a

Hinze, S., et al. (2019) 'Different Processes, Similar Results? A Comparison of Performance Assessment in Three Countries'. In Springer Handbook of Science and Technology Indicators (pp. 465-484). Springer, Cham.

Ibarra Unzueta, J. A., Castro Spila, J., and Barrenechea, J. I. (2007) '*La evaluación de la actividad científica en Ciencias Sociales y Humanidades'*. Editorial de la Universidad del País Vasco.

Jamali, H. R., et al. (2020) 'National comparisons of early career researchers' scholarly communication attitudes and behaviours', *Learned Publishing*, 33/4: 370-384. doi: 10.1002/leap.1313

Jiménez-Contreras, E., de Moya Anegón, F., and Delgado-López-Cózar, E. (2003) 'The evolution of research activity in Spain: The impact of the National Commission for the Evaluation of Research Activity (CNEAI)', *Research policy*, 32/1: 123-142. doi: 10.1016/S0048-7333(02)00008-2.

Jiménez-Contreras, E. et al. (2002) 'Impact-factor rewards affect Spanish research' *Nature*, 417/6892: 898. doi: 10.1038/417898b

John, L. K., Loewenstein, G., and Prelec, D. (2012) 'Measuring the prevalence of questionable research practices with incentives for truth telling', *Psychological science*, 23/5: 524-532. doi: 10.2139/ssrn.1996631

Laudel, G., and Gläser, J. (2006) 'Tensions between evaluations and communication practices', *Journal of Higher Education Policy and Management*, 28/3: 289-295. doi: 10.1080/13600800600980130

Levecque, K., et al. (2017) 'Work organization and mental health problems in PhD students', *Research Policy*. 46: 868-879. doi: 10.1016/j.respol.2017.02.008

Maggio, L., et al. (2019) 'Factors associated with scientific misconduct and questionable research practices in health professions education', *Perspectives on medical education*, 8/2: 74-82. doi: 10.1007/s40037-019-0501-x

Marini, G. (2018) 'Tools of individual evaluation and prestige recognition in Spain: how sexenio 'mints the golden coin of authority', *European journal of higher education*, 8/2: 201-214. doi: 10.1080/21568235.2018.1428649

Ministerio de Educación, Cultura y Deporte. (2014) Datos y Cifras del Sistema Universitario Español. Curso 2013–2014. Madrid http://www.educacionyfp.gob.es/dam/jcr:0878b70a-d7d3-48ff-9587-9177f6dd57db/datos-cifras-13-14.pdf

Ministerio de Universidades. (2020) Datos y Cifras del Sistema Universitario Español. Publicación 2019–2020. Madrid https://www.ciencia.gob.es/stfls/MICINN/Universidades/Ficheros/Estadisticas/Informe_Datos_Cifras_Sistema_Universitario_Espanol_2019-2020.pdf

Molas-Gallart, J. (2012) 'Research Governance and the Role of Evaluation. A Comparative Study' American Journal of Evaluation 33/4: 583–598. doi: 10.1177/1098214012450938

Murphy, S. (2017) *Zombie university: Thinking under control*. Watkins Media Limited.

Narin, F. (1976) *Evaluative bibliometrics: The use of publication and citation analysis in the evaluation of scientific activity*. New Jersey: Computer Horizons.

Nederhof, A. J. (2006) 'Bibliometric monitoring of research performance in the social sciences and the humanities: A review', *Scientometrics*, 66/1: 81-100. doi: 10.1007/s11192-006-0007-2

Nicholas, D., et al. (2020a). Does the scholarly communication system satisfy the beliefs and aspirations of new researchers? Summarizing the Harbingers research. *Learned Publishing*, *33*/2: 132-141. doi: 10.1002/leap.1284

Nicholas, D., et al. (2020b) 'Millennial researchers in a metric-driven scholarly world: An international study', *Research Evaluation*, 1–12. doi: 10.1093/reseval/rvaa004

Ochsner, M., Hug, S. E., and Daniel, H.-D. (2012) 'Four Types of Research in the Humanities: Setting the Stage for Research Quality Criteria in the Humanities', *Research Evaluation*, 22/2: 79–92. doi: 10.1093/reseval/rvs039

Ossenblok, T. L., Engels, T. C., and Sivertsen, G. (2012) 'The representation of the social sciences and humanities in the Web of Science—a comparison of publication patterns and incentive structures in Flanders and Norway (2005–9)', *Research Evaluation*, 21/4: 280-290. doi: 10.1093/reseval/rvs019

Osuna, C., Cruz-Castro, L., and Sanz-Menéndez, L. (2011). Overturning some assumptions about the effects of evaluation systems on publication performance. *Scientometrics*, 86/3): 575-592.  doi: 10.1007/s11192-010-0312-7.

Pupovac, V., Prijić-Samaržija, S., and Petrovečki, M. (2017) 'Research misconduct in the Croatian scientific community: a survey assessing the forms and characteristics of research misconduct'. *Science and engineering ethics*, 23/1: 165-181. doi: 10.1007/s11948-016-9767-0

Rey-Rocha, J., et. al (2001) 'Some misuses of journal impact factor in research evaluation' *Cortex*, 37: 595–597. doi: 10.1016/s0010-9452(08)70603-8

Rijcke, S. D., et al. (2016). Evaluation practices and effects of indicator use—a literature review. Research Evaluation, 25(2), 161-169.  doi: 10.1093/reseval/rvv038.

Rodríguez, A., and Soldevilla, S. G. (eds.). (2018) *Investigar en la era neoliberal: visiones críticas sobre la investigación en comunicación en España*, Valencia: Universitat de València.

Rodríguez-Bravo, B., and Nicholas, D. (2020) 'Descubrir, leer, publicar, compartir y monitorizar el progreso: comportamiento de los investigadores junior españoles', *Profesional de la información*, 29/5: 1-16. doi: 10.3145/epi.2020.sep.03

Rodríguez-Bravo, B., and Nicholas, D. (2018) Reputación y comunicación científica: investigadores españoles en el inicio de su carrera. El profesional de la información, 28/2,  doi: 10.3145/epi.2019.mar.03

Rodríguez-Yunta, L. (2017) 'Evaluación de publicaciones en humanidades: cambios necesarios frente a indicadores inadecuados'. *Anuario ThinkEPI*, 11: 230-240. doi: 10.3145/thinkepi.2017.43

Rovira, L. (2006) ¿Hacia una evaluación métrica de la investigación en las humanidades y en las ciencias sociales?. *La evaluación de la actividad científica en Ciencias Sociales y Humanidades*, 31-52. http://hdl.handle.net/10810/26180

Ruiz-Pérez, R., Delgado-López-Cózar, E., and Jiménez-Contreras, E. (2010) 'Principios y criterios utilizados en España por la Comisión Nacional Evaluadora de la Actividad Investigadora (CNEAI) para la valoración de las publicaciones científicas: 1989-2009', *Psicothema*, *22*/4: 898-908.

Sanz-Menéndez, L. (1995) 'Research actors and the state: research evaluation and evaluation of science and technology policies in Spain', *Research Evaluation*, 5(1): 79-88. doi: 10.1093/rev/5.1.79

Sanz-Menéndez, L., and Cruz-Castro, L. (2019) 'University academics' preferences for hiring and promotion systems', *European Journal of Higher Education*, 9/2: 153-171. doi: 10.1080/21568235.2018.1515029

Urbano, C., Borrego, À. B. H.,. and Rodríguez-Gairín, J. (2004) 'La evaluación de revistas en procesos de evaluación de la investigación española en humanidades y ciencias sociales: desarrollo de un modelo y experimentación en el área de filología hispánica'. *Informe Programa de Estudios y Análisis del Ministerio de Educación y Ciencia*. Madrid http://diposit.ub.edu/dspace/bitstream/2445/33405/1/ea0025_evaluacion_revistas_proce sos_evaluacion.pdf

Van Den Besselaar, P., Heyman, U., and Sandström, U. (2017) 'Perverse effects of output-based research funding? Butler's Australian case revisited', *Journal of Informetrics*, *11*/3: 905-918. doi: 10.1016/j.joi.2017.05.016

Whitley, R. (2007). Changing governance of the public sciences. In editors: Whitley, R. and Gläser, J. (eds.), *The changing governance of the sciences* (pp. 3-27). Springer: Dordrecht.

Wilsdon,J., et al. (2015) 'The Metric Tide: Report of the Independent Review of the Role of Metrics in Research Assessment and Management'. Bristol, UK. doi: 10.13140/RG.2.1.4929.1363

Wouters, P. (2014) 'The Citation: From Culture to Infrastructure'. In Cronin B and Sugimoto C. R. (eds). *Beyond Bibliometrics: Harnessing Multidimensional Indicators of Scholarly Impact* (pp. 47–66). Cambridge MA: MIT press.

Wouters, P., et al. (2015) 'The Metric Tide: Literature Review'. *Supplementary Report I to the Independent Review of the Role of Metrics in Research Assessment and Management*. Bristol, UK. HEFCE. doi: 10.13140/RG.2.1.5066.3520.